# Whale Optimization Algorithm-based Fractional Order Fuzzy Type-II PI Control for Modular Multilevel Converters

Mohammad Ali Labbaf-Khaniki[1], Mohammad Manthouri[2], Amin Hajizadeh[3]

Faculty of Electrical Engineering, K.N. Toosi University of Technology, Tehran, Iran[1]

Department of Electrical and Electronic Engineering, Shahed University, Tehran, Iran.[2]

Department of Energy Technology, Aalborg University, Esbjerg, Denmark.[3]

mohamad95labafkh@gmail.com[1], mmanthouri@shahed.ac.ir[2], aha@et.aau.dk[3]

**Abstract:** Designing a robust controller for Modular Multilevel Converters (MMCs) is crucial to ensure stability and optimal dynamic performance under various operating conditions, including faulty and disturbed scenarios. The primary objective of controlling grid-connected MMCs (GC-MMCs) is to accurately track real and reactive power references while maintaining excellent harmonic performance in the output response. This paper proposes a novel model-free control strategy for GC-MMCs, employing a Fractional Order Proportional-Integral (FOPI) controller and a Fractional Order Fuzzy type-II Proportional-Integral (FOFPI) controller. The FOFPI controller utilizes a type-II Fuzzy Inference System (FIS) to adaptively adjust the proportional and derivative gains during the control process, enabling effective control of the MMC under diverse operating conditions. The type-II FIS, which leverages type-II fuzzy sets, can mitigate uncertainty and nonlinearity in the system. Furthermore, the incorporation of fractional-order mathematics enhances the flexibility of the proposed controllers. To optimize the initial parameters of the proposed controllers, the Whale Optimization Algorithm (WOA), a meta-heuristic algorithm, is employed. The results demonstrate that the proposed controllers exhibit superior performance under voltage disturbance conditions, varying input voltage, and can ensure the stability of the MMC.

**Keywords.** Modular Multilevel Converter, Fractional Order PI Controller, Type-II Fuzzy Inference System, Whale Optimization Algorithm.

## 1) Introduction

Modular multilevel converters (MMCs) have emerged as a promising solution for providing high reliability, capability, and reasonable harmonic performance in renewable power generation systems [1]. Due to their high-power handling capabilities, MMCs are commonly employed in grid-connected converters [2]. These converters consist of a large number of capacitors and power electronic switches. To maintain a constant DC voltage across the capacitors and compensate for power losses, the MMC absorbs a small amount of active power. However, mismatches in conduction and switching losses of the power electronic switches can lead to unbalanced capacitor voltages [3]. Balancing the submodule (SM) capacitor voltages has been a significant technical challenge in recent research [4] and [5]. To date, numerous papers have been published on voltage balancing control methods, which can be categorized into two groups: distributed voltage closed-loop control methods and centralized SM selection control methods [6]. Both approaches involve controlling capacitor voltages by regulating the insulated-gate bipolar transistors (IGBTs) during the controlled precharge and normal operation stages of the high-voltage direct current (HVDC) system. [7] This paper proposes a hierarchical control strategy to enhance the voltage profile in islanded microgrids. The strategy consists of three control layers: inner, middle, and outer layers. The inner layer controls individual MMCs, the middle layer coordinates the operation of multiple MMCs, and the outer layer oversees the entire microgrid. Furthermore, investigations have been conducted on the implementation of advanced current controllers for MMCs. [8] proposed an MPC control strategy based on group sorting, which reduces the number of optimization times by half. This method divides all SMs in the arm into several groups, and the SMs within each group are settled, although this approach cannot guarantee optimal capacitance-voltage balance control outcomes. In [9], the control and modeling of MMCs with grid-connected photovoltaic systems are also presented. One of the most significant technical challenges in MMCs is the synchronized control of variables, including AC current and voltage, circulating currents, and capacitor voltages of the arms. As switching converters, MMCs exhibit nonlinear behavior due to their numerous control variables and nonlinear dynamic equations. Therefore, advanced and nonlinear methods superior to classical linear techniques are required. Conventional linear-based controllers, such as PI and PR, often rely on complex cascade or parallel structures, decoupling assumptions between control variables, and tuning multiple MMC parameters [10].

Synchronization is a critical issue in grid-connected systems. Although conventional proportional-integral (PI) controllers used for synchronization provide adequate performance, they yield poor transient response due to poor tuning of controller gain values [11] and [12]. To overcome this limitation, this study designs and implements a fuzzy-tuned PI controller. The fuzzy logic controller tunes the controller gains based on the operating point of the system [13]. The primary advantage of using fuzzy-tuned PI is to leverage the benefits of both PI and fuzzy logic controllers (FLCs) [14]. The resultant response overcomes the sluggish response and reduced steady-state error of FLCs, while maintaining the advantages of low overshoot and settling time [15], [16]. Due to their simplicity, PID controllers are widely used in many industrial systems and applications. One effective method for controlling Grid-Connected MMCs (GC-MMCs) is using PI current controllers. However, the presence of noise in the design makes the use of the controller's derivative unsuitable due to noise amplification. In classical PID controllers, the order of the derivative and integral parts is considered as one, whereas in fractional-order PID (FOPID) controllers, the order of the derivative and integral is set between 0 and 2 ($PI^\alpha D^\beta$). Furthermore, combining intelligent controllers, such as fuzzy controllers, with classical controllers can help control the system in uncertainty and nonlinearity[17] and [18].

Fuzzy logic, a mathematical approach to handle uncertainty and imprecision, is employed in the proposed controller to effectively manage the nonlinear and uncertain behavior of the MMC. Also, fuzzy logic has been used in many applications. This paper presents a novel approach to image noise reduction using stochastic computing-based fuzzy filtering. The proposed method achieves efficient noise reduction while reducing computational cost [19]. This paper [20] proposes a novel information-theoretic approach to testing and debugging fairness defects in deep neural networks. The authors introduce a framework that leverages mutual information and conditional entropy to identify and quantify biases in neural networks. Type-II Fuzzy Inference Systems (FIS) are more effective in withstanding uncertainty due to their structure and membership functions, compared to Type-I FIS [21]. The CTLBO approach is a novel optimization technique designed to tackle dynamic multi-objective problems, which involve optimizing multiple conflicting objectives that change over time [22].

This paper proposes a novel control strategy for MMCs based on a Fractional Order Fuzzy Type-II PI Controller, which is optimized using the Whale Optimization Algorithm to achieve improved performance and robustness in the presence of uncertainties and disturbances. Here are the contributions of the paper as bulleted points:

- Design of a Fractional-Order Proportional-Integral (FOPI) current controller to control GC-MMCs under unbalanced voltage conditions and model uncertainties.
- Development of a Type-II FOFPI current controller to improve the control performance of GC-MMCs. Tuning of the controller's coefficients, including the gains of proportional and integral terms, and the order of integral in the FOPI controller, using WOA. Tuning of the parameters of MFs in the FOFPI controller using WOA.

- Simulation study to validate the effectiveness of the proposed control strategy under normal and unbalanced fault conditions. Presentation of the obtained results to demonstrate the potential of the recommended control strategy for GC-MMCs.

This paper is organized as follows: The modeling of GC-MMCs is presented in Section II. Section III describes the proposed control scheme, while Section IV provides a detailed presentation of the simulation results.

## 2) Modelling of Grid Connected Modular Multilevel Converter

GC-MMCs are a type of power electronic converter that have gained popularity in recent years due to their ability to efficiently convert AC power from the grid to DC power for various applications. GC-MMCs are composed of multiple identical submodules, each consisting of power electronic switches, diodes, and capacitors, which are connected in series to form a modular structure. This modular design enables GC-MMCs to achieve high voltage and power ratings, making them suitable for high-power applications such as renewable energy systems, high-voltage direct current (HVDC) transmission, and medium-voltage drives. The advantages of GC-MMCs include their high scalability, flexibility, and reliability, as well as their ability to provide high-quality output waveforms and low harmonic distortion. However, the control of GC-MMCs is complex due to the large number of submodules and the nonlinear behavior of the system, requiring advanced control strategies to ensure stable and efficient operation.

To evaluate the dynamic performance of GC-MMCs, a precise model is essential. The classic configuration of an MMC is illustrated in Fig. 1, where each SM consists of a simple half-bridge formed by two power electronic switches, two anti-parallel diodes, and a capacitor C. To develop an advanced control strategy for GC-MMCs, nonlinear state-space equations based on the average model of power electronic converters are employed. According to Fig. 1, the differential equations of an N-cells MMC can be derived using basic Kirchhoff's Current Law (KCL) and Kirchhoff's Voltage Law (KVL):

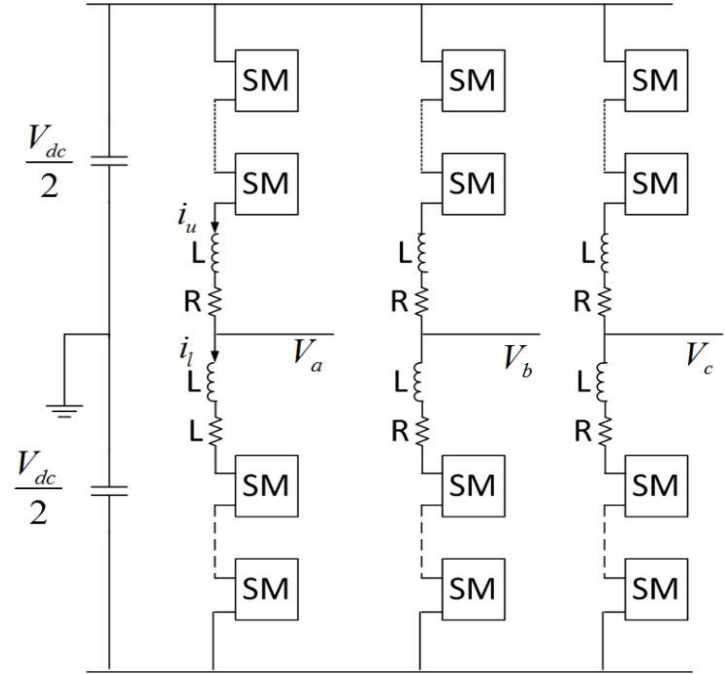

**Fig.1.** The circuit topology of a typical MMC [8]

$$\frac{di_u}{dt} = \frac{1}{L}\left[\frac{V_{dc}}{2} - \sum_{i=1}^{N}(d_i.V_{C_{i-u}}) - Ri_u - V_a\right] \tag{1}$$

$$\frac{di_l}{dt} = \frac{1}{L}\left[\frac{V_{dc}}{2} - \sum_{i=N+1}^{2N}(d_i.V_{C_i-l}) - Ri_l + V_a\right] \tag{2}$$

$$\frac{dV_{C_i-u}}{dt} = \frac{1}{C}(i_u.d_i) \qquad i = 1, \dots, N \tag{3}$$

$$\frac{dV_{C_i-l}}{dt} = \frac{1}{C}(i_l.(1-d_i)) \qquad i = N+1, \dots, 2N \tag{4}$$

The variables $i_u$, $i_l$ represent the upper and lower arm currents, respectively, while $V_{C_i-u}$, $V_{C_i-l}$ denote the upper and lower capacitor voltages, respectively. Additionally, $d_i$ represents the gating signal of the upper gate of the i-th cell, $V_{dc}$ is the DC-link voltage, and $V_a$ is the phase a voltage.

### 3) Controller Description

This article employs the vector control method, also known as field-oriented control, to regulate and minimize the Total Harmonic Distortion (THD) of the output voltage. To achieve this, the difference between the reference and actual values of the direct-axis current ($i_d$) and quadrature-axis current ($i_q$) is defined as an error signal. Based on this approach, two controllers are proposed: the FOPI controller and the FOFPI controller. Each controller is designed to reduce the corresponding error, thereby ensuring precise control of the output voltage.

### 3.1) FOPI controller

The classical PI controller is a widely used control strategy in process control and automation. It consists of two components:

- Proportional (P) term: This term is proportional to the error signal, which is the difference between the desired setpoint and the actual process variable. The proportional gain, $K_P$, determines the magnitude of the correction signal.
- Integral (I) term: This term is proportional to the integral of the error signal over time. The integral gain, $K_I$, determines the magnitude of the correction signal.

The PI controller combines these two terms to produce a control signal that adjusts the process variable to achieve the desired setpoint. In the Laplace domain, the transfer function of a PI controller can be represented as:

$$u_{PI}(s) = (K_P + \frac{K_I}{S}).E(s) \qquad (5)$$

Where $u_{PI}(s)$ is the control signal in the Laplace domain, $E(s)$ is the error signal in the Laplace domain, $K_P$ is the proportional gain, $K_I$ is the integral gain, $s$ is the Laplace variable. In the time domain, the control signal can be represented as:

$$u_{PI}(t) = K_P e(t) + K_I \int_0^t |e(t)|\, dt \qquad (6)$$

where $u_{PI}(t)$ is the control signal in the time domain, $e(t)$ is the error signal in the time domain, $K_P$ is the proportional gain, $K_I$ is the integral gain, $\int_0^t |e(t)|\, dt$ is the integral of the error signal over time. The PI controller is widely used because it is simple to implement and can provide good control performance for many systems. However, it has some limitations, such as:

- It can be sensitive to changes in the process dynamics
- It may not provide good disturbance rejection
- It can be difficult to tune the gains for optimal performance

To overcome these limitations, more advanced control strategies, such as fractional-order PI controllers and fuzzy PI controllers, have been developed. These controllers can provide more robust and adaptive control performance, but they are also more complex and require more sophisticated tuning methods.

Several approaches have been proposed to simulate fractional-order calculations. One of the most popular methods is the Oustaloup recursive approximation, which approximates fractional integrals and derivatives using a linear filter with an order of $2N + 1$ [23]. This method is effective within a specific frequency range, namely $[\omega_L, \omega_H]$, as described in [24].

$$G_f(s) = S^\alpha = K \prod_{k=-N}^{N} \frac{S + \omega'_k}{S + \omega_k} \qquad (7)$$

where $\omega_k$ is the zeroes of the filter, $\omega'_k$ is the poles of the filter and $K$ is the gain of the it.

$$\omega'_k = \omega_b \left(\frac{\omega_h}{\omega_b}\right)^{\frac{k+N+0.5*(1-\alpha)}{2N+1}} \qquad (8)$$

$$\omega_k = \omega_b \left(\frac{\omega_h}{\omega_b}\right)^{\frac{k+N+0.5*(1+\alpha)}{2N+1}} \qquad (9)$$

$$K = \omega_h^\alpha \qquad (10)$$

The accuracy of the Oustaloup recursive approximation is directly influenced by the order of the filter ($N$). As $N$ increases, the approximation becomes more precise, but also more complex. The frequency range of the approximation must be carefully selected to match the operating frequency of the controlled system. In this study, a filter order of $N = 20$ is chosen, with a frequency range of [0.001, 1000] rad/s. Notably, the performance of the FOPI controller is highly sensitive to the value of $\alpha$, which plays a critical role in determining its effectiveness.

### 3.2) Fuzzy inference system

Fuzzy logic is a computational approach that mimics human thinking and reasoning, which is why fuzzy systems are categorized as intelligent systems. At the core of fuzzy logic lies a set of IF-Then rules, which are structured as an inference system known as a Fuzzy Inference System (FIS). An FIS typically consists of three main components:

- Fuzzifier: This component converts crisp input values into fuzzy sets, which are linguistic variables with membership degrees.
- Inference Engine: This component applies the IF-Then rules to the fuzzified inputs, using logical operators such as AND, OR, and NOT to combine the rules.

- Defuzzifier: This component converts the fuzzy output of the inference engine into a crisp value, which is the final output of the FIS.

The output of the FIS is a crisp value that represents the conclusion drawn from the IF-Then rules. This output can be used to control a system, make decisions, or classify patterns. In the context of control systems, FIS is often used to tune the parameters of a controller, such as the gains of a PI controller. The FIS takes the error signal and its derivative as inputs, and outputs the adjusted gain values that minimize the error. There are two main types of FIS:

- Type-I FIS: This type of FIS uses a simple and intuitive approach to fuzzy inference, where the output of the FIS is a weighted sum of the outputs of each rule.
- Type-II FIS: This type of FIS is more advanced and uses a more sophisticated approach to fuzzy inference, where the output of the FIS is a fuzzy set that represents the uncertainty of the output.

In the context of the paper, the FOFPI controller uses a Type-II FIS to tune the parameters of the PI controller, which allows for more robust and adaptive control of the system. The output of the FIS is

$$y(x) = \frac{\sum_{j=1}^{m} y^j \left( \prod_{i=1}^{n} \mu_{A_i^j}(x_i) \right)}{\sum_{j=1}^{m} \prod_{i=1}^{n} \mu_{A_i^j}(x_i)}, \qquad (11)$$

where $\mu_{A_i^j}$ is the membership function grade for the input $x_i$, which is also referred to as the antecedent part of the FIS. $y^j$ is the corresponding output for each rule, which is also referred to as the consequent part of the FIS. $n$ is the number of inputs to the FIS. $m$ is the number of rules in the FIS. The normalized product of the grades in the antecedent part can be represented as:

$$\xi^j(x) = \frac{\prod_{i=1}^{n} \mu_{A_i^j}(x_i)}{\sum_{j=1}^{m} \prod_{i=1}^{n} \mu_{A_i^j}(x_i)}, \tag{12}$$

This equation calculates the weighted product of the membership function grades for each rule, which is then normalized by the sum of the products for all rules.

$$y(x) = \theta^T \xi(x), \tag{13}$$

where $\theta$ is is the center of the consequent part of MFs ($\theta = (y^1, \dots, y^m)^T$) and ($\xi(x) = (\xi^1(x), \dots, \xi^m(x))^T$). Based on above, the type-II FIS can be defined as follows

$$y_{T-II}(x) = m\theta^T \xi_U(x) + (1-m)\theta^T \xi_L(x) \tag{14}$$

The normalized product of the grades in lower and upper type-II membership functions (MFs) is denoted by $\xi_U(x)$ and $\xi_U(x)$, respectively. The constant m determines the effect ratio of the lower and upper type-II MFs. In other words, $\xi_U(x)$ and $\xi_U(x)$ represent the normalized products of the grades in the upper and lower type-II membership functions, respectively, with m being a constant that influences the relative impact of the lower and upper type-II MFs. The paper discusses a Fractional Order Fuzzy Proportional Integral (FOFPI) controller, where the error ($e$) and derivative of error ($\dot{e}$) serve as inputs, and the proportional and integral coefficients ($K_P, K_I$) of the FOFPI controller are the outputs. The Fuzzy Inference System (FIS) output is shown

$$y_{T-II}\big((e,\dot{e})|\theta_{K_P}\big) = m\theta_{K_P}^T \xi_U(e,\dot{e}) + (1-m)\theta_{K_P}^T \xi_L(e,\dot{e}), \tag{15}$$

$$y_{T-II}\big((e,\dot{e})|\theta_{K_I}\big) = m\theta_{K_I}^T \xi_U(e,\dot{e}) + (1-m)\theta_{K_I}^T \xi_L(e,\dot{e}), \tag{16}$$

Where $y_{T-II}\big((e,\dot{e})|\theta_{K_P}\big)$ and $y_{T-II}\big((e,\dot{e})|\theta_{K_I}\big)$ are the FIS outputs and generates the gains of FOFPI controller, $\theta_{K_P}^T$ and $\theta_{K_I}^T$ are respectively, the centers of MFs in consequent part for $K_P$ and $K_I$. With respect to (15-16), the FOFPI controller formula can be computed in (17).

$$U_{FOFPI}(t) = y_{T-II}\big((e,\dot{e})|\theta_{K_P}\big)e(t) + y_{T-II}\big((e,\dot{e})|\theta_{K_I}\big)\,_0I_t^\alpha e(t) \tag{17}$$

The FOFPI controller's control signal, $U_{FOFPI}(t)$, is designed to adapt to changing conditions during the control process. This adaptability is achieved through varying gains, which enables the FOFPI controller to outperform the FOPI controller with fixed MFs used in the antecedent part of the FIS and the controller's structure are illustrated in Figures 2 and 3, respectively.

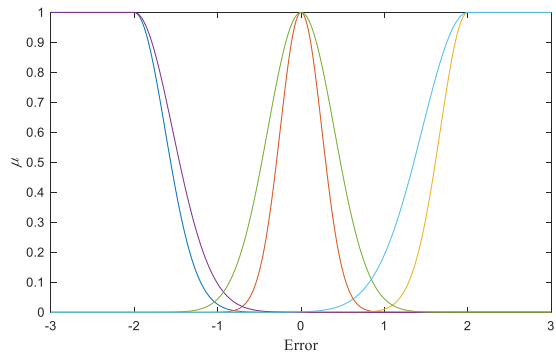

Fig. 2(a). Type-II MFs of the first input (error)

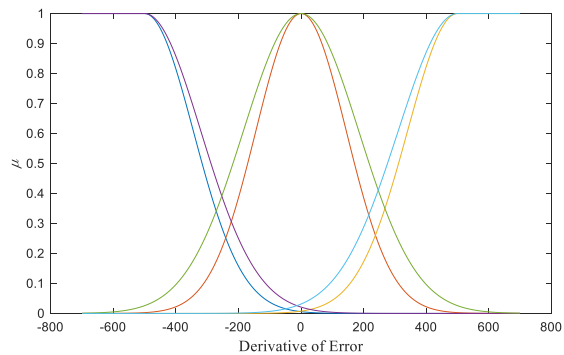

Fig. 2(b). Type-II MFs of the second input (derivative of error)

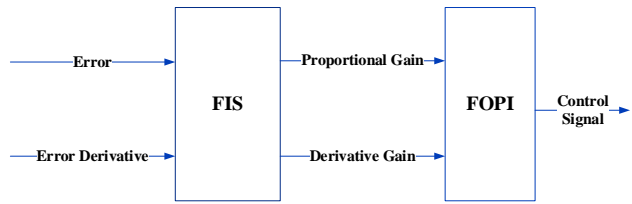

Fig.3. Structure of the FOFPI controller

### 3.3) Whale Optimization Algorithm

The Whale Optimization Algorithm (WOA) is a bio-inspired algorithm that mimics the hunting behaviour of humpback whales. The algorithm consists of two stages: exploration and exploitation. In the exploration phase, whales move in a spiral path around their prey, simulating the hunting behaviour of humpback whales. In the exploitation phase, whales use a mechanism called bubble-net feeding, where they create distinctive bubbles along a spiral to surround and capture their prey. The algorithm updates the position of whales using two strategies: shrinking encircling and spiral updating position, with a probability of 50% for each strategy. The mathematical model of the algorithm is defined using equations that simulate the behaviour of whales. The algorithm has been used to solve various optimization problems and has been shown to be effective in finding low-quality solutions. However, it suffers from poor search strategies, resulting in low exploitation, exploration, and local optima trapping.

After generating a set of random populations, the best solution is identified as the leading search agent, and the remaining populations adjust their positions to converge towards it [25]. The primary formula is presented as follows

$$X(t+1) = \begin{cases} D'e^{bl}\cos(2\pi t) + X^*(t) & p \geq 0.5 \\ X^*(t) - AD & p < 0.5 \end{cases} \quad (18)$$

The equation describes a search algorithm that uses a logarithmic spiral to explore a space. The algorithm is based on a random number $p$ between 0 and 1, a random number l between -1 and 1, and a shape parameter $b$ that determines the spiral's shape. The algorithm also uses the current iteration $t$ and the best position found so far $X^*$. (20-21) represents $D'$ and $D$.

$$D' = |X^*(t) - X(t)| \quad (19)$$

$$D = |CX^*(t) - X(t)| \quad (20)$$

where $X(t)$ is the position of the populations, A and C are constant and it is given by:

$$A = 2ar - a \tag{21}$$

$$C = 2r \tag{22}$$

The equation involves two key components: a decreasing coefficient $a$ that ranges from 2 to 0 over each iteration, and a random vector $r$ that falls between 0 and 1. The first part of the equation models the encircling behavior of humpback whales as they surround their prey, while the second part simulates the unique bubble-net attack strategy used by these whales to catch their prey.

The optimization problem is formulated to minimize the Total Harmonic Distortion (THD) of the line-to-line voltage. The WOA is employed to optimize the parameters of two controllers: the FOPI controller and the FOFPI controller. Specifically, WOA tunes the three coefficients of the $(PI^\alpha)$ controller, including the proportional, integral, and order of integral terms, as well as the center and sigma values of the MFs in the antecedent and consequent parts of the FIS in the FOFPI controller.

**4) Simulations**

This study examines the performance of two proposed controllers under different input voltage conditions. Specifically, the validation of the controllers is tested using two distinct input voltage levels: 400 volts and 600 volts. The results are visualized in Figure 4, which displays the three-phase line-line voltage waveform.

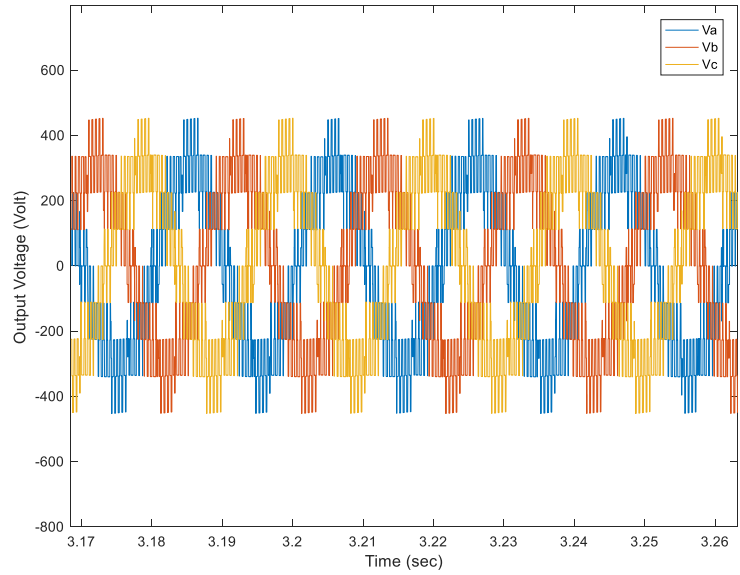

(a)

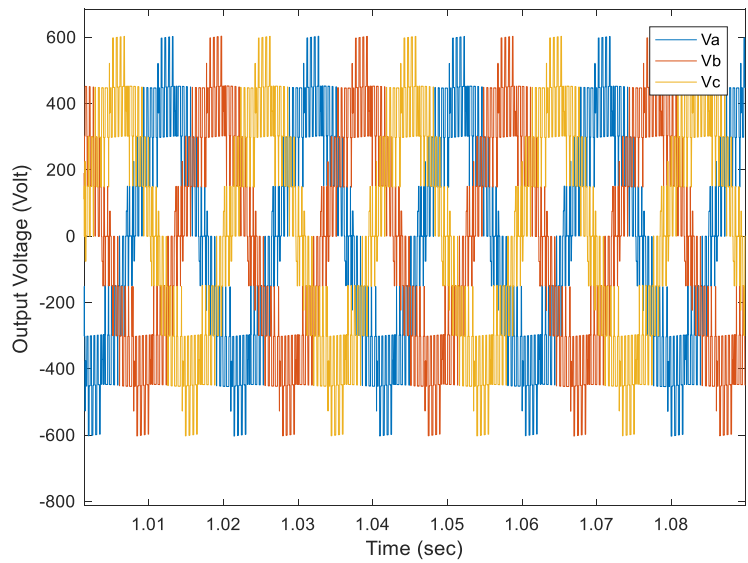

(b)

Fig. 4(a). three-phase line-line voltage (Input=450) and (b). three-phase line-line voltage (Input=600)

The results show that the voltage levels remained within acceptable limits throughout the charging and discharging cycles of the capacitors. Furthermore, as shown in Table (1), the THD of the output voltage significantly decreased when the FOFPI controller was implemented, indicating a notable improvement in voltage quality.

Table. (1). THD of the output voltage using the FOPI and FOFPI controllers.

|  | FOPI | FOFPI |
|---|---|---|
| 500 volts | 0.31 | 0.26 |
| 400 volts | 0.33 | 0.28 |

Figure 5 illustrates the dynamic adjustments made to the FOFPI controller's coefficients, $K_P, K_I$, as it regulates the $i_d, i_q$ currents. The scenario involves a step change in input voltage, from 450 volts to 600 volts, at time $t = 1s$, allowing for a detailed examination of the controller's performance under varying conditions. The voltage levels remained stable and within acceptable limits throughout the entire process of capacitor charging and discharging.

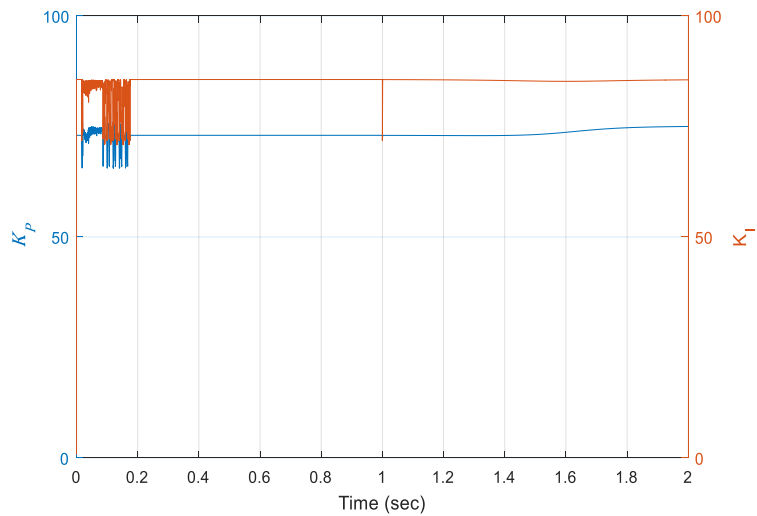

Fig. 5(a). Changes of the parameters of FOFPI controller to control $i_d$

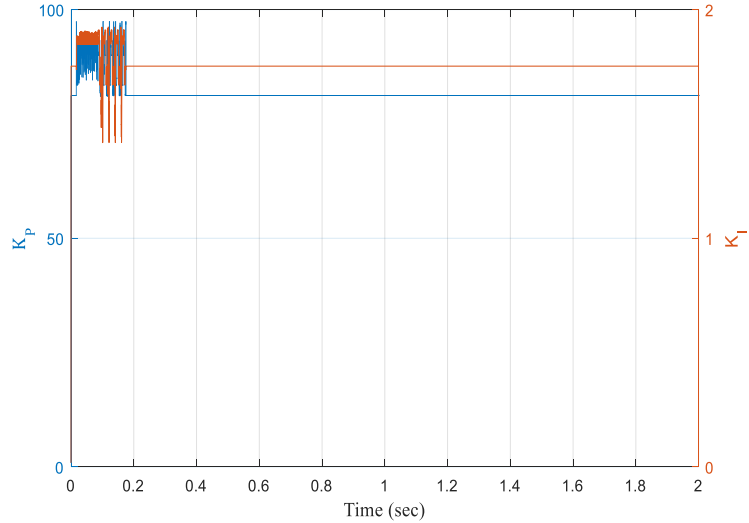

Fig. 5(b). Changes of the parameters of FOFPI controller to control $i_q$

As shown in Figure 5, the FOFPI controller adapts to changes in the system by adjusting its coefficients. During the transient state, the proportional gain ($K_p$) is increased to accelerate the system's response, while the $K_I$ is decreased. In contrast, during the steady state, $K_p$ remains constant, and $K_I$ is increased to minimize the steady-state error.

5) **Conclusion**

This paper presents a robust vector control approach for Grid-Connected Modular Multilevel Converters (GC-MMCs). Two controllers, FOFPI type-II and FOPI, are designed to tackle the nonlinear and uncertain nature of the GC-MMC plant. The results show that the FOFPI controller outperforms the FOPI controller, achieving a smoother output line-line voltage with reduced ripple and THD. The FOFPI controller's adaptability to changing gains in response to faults and varying operating conditions enables it to surpass the FOPI controller's performance. Furthermore, the WOA successfully optimized the parameters of both controllers. The proposed controller ensures system stability across different operating conditions, even in the presence of uncertainties and faults.

The proposed FOFPI controller can be extended to other converter topologies, such as cascaded H-bridge converters or modular multilevel converters with different configurations. Also, we can compare the performance of the proposed FOFPI controller with other advanced control strategies, such as Model Predictive Control (MPC) or Sliding Mode Control (SMC), to evaluate its advantages and limitations.